# Имитационное моделирование субмикросекундных режимов работы высоковольтных реверсивно-включаемых динисторов

*А.В. Горбатюк*,[1,2] *Б.В. Иванов*[3]

[1] Физико-технический институт им. А.Ф. Иоффе РАН, Санкт-Петербург, Россия
e-mail: agor.pulse@mail.ioffe.ru

[2] Санкт-Петербургский государственный политехнический университет, Санкт-Петербург, Россия

[3] Санкт-Петербургский государственный электротехнический университет "ЛЭТИ" им. В.И.Ульянова (Ленина), Санкт-Петербург, Россия

На основе 2-мерного компьютерного моделирования физических процессов в высоковольтных реверсивно-включаемых динисторах (РВД) изучены важные особенности процессов коммутации ими субмикросекундных импульсов тока. Характерно, что на фронтах этих импульсов наблюдаются большие выбросы напряжения и опасные задержки нарастания тока. Установлено, что эти нежелательные проявления могут быть устранены при подходящем ослаблении легирования $p$-базы. Следуя этому приему для РВД с напряжением переключения $2.5 - 5$ kV можно уменьшить длительность фронтов включения до $75 - 150$ ns и ниже при величине заряда реверсивной накачки $5 - 10$ $\mu C/cm^2$. Скорость нарастания тока при этом может быть увеличена до $(2 - 3)10^{10}$ $A \cdot cm^{-2} s^{-1}$, то есть, более чем на порядок выше этого показателя для стандартных РВД. При амплитудах коммутируемых импульсов в единицы kA и длительностях в несколько сотен наносекунд передаваемая в нагрузку энергия может достигать долей и единиц J за импульс при потерях на нагрев РВД около 10% от этих значений.

## Введение

Высоковольтные переключатели тиристорного типа с повышенным быстродействием остро востребованы в современной импульсной энергетике [1]. В качестве перспективных кандидатов рассматриваются реверсивно-включаемые динисторы (РВД), зарекомендовавшие себя как самые мощные и надежные переключатели субмилли- и микросекундного диапазонов [2 – 7]. Однако, хотя принципиальная возможность включения субмикросекундных импульсов тока на основе РВД была обнаружена еще в лабораторных экспериментах [8], они до сих пор остаются непригодными для практического использования в мощных коммутаторах из-за ряда физических ограничений и противоречивых технических требований.

Возникающие проблемы были частично исследованы в нашей расчетной работе [9]. Выяснилось, что при функционировании РВД в быстрых коммутационных режимах с недостаточно мощным управлением наблюдается появление переходных всплесков напряжения ~ 1 kV и задержек нарастания тока длительностью в десятки ns при достижении им плотностей в сотни $A/cm^2$. Это может вызвать не только рост интегральных потерь, но и аварийные локализации тока и тепла. Известно, что ослабить подобные проявления в РВД стандартных конструкций можно при сильном увеличении амплитудной плотности управляющих импульсов. Так, в экспериментах [8] она доходила до 1 $kA/cm^2$. Но на практике это ведет к заметному снижению к.п.д. коммутатора в целом из-за дополнительных потерь в мощном управляющем генераторе. Да и сама разработка высоковольтных наносекундных генераторов на такие токи является далеко не простой задачей [4]. В работе [9] расчетным путем было показано, что задержку можно сократить до единиц наносекунд, если уменьшить толщину $p$-базы катодного транзистора от типичных значений $50 -$ 80 μm до величин, ниже 10 μm. Тем не менее, исключить ее полностью не удается, и без дополнительных исследований этот прием нельзя принимать за окончательное решение.

В настоящей работе продолжается изучение возможностей РВД. В ней проводятся детальные компьютерные эксперименты по субмикросекундным режимам функционирования РВД при учете его взаимодействия с цепью нагрузки для нескольких наборов параметров конструкции и элементов цепей и вырабатываются практические рекомендации, направленные на значительное повышение быстродействия высоковольтных коммутаторов на РВД.

### Особенности субмикросекундного РВД и методика его моделирования

Напомним базисные принципы РВД [1 – 3]. Их кремниевая структура состоит из большого числа ($k = 10^2-10^3$) одинаковых элементарных ячеек (ЭЯ), в каждой из которых можно выделить тиристорные ($p^+n'n_0pp'n^+$) и диодные ($n'n_0pp'$) элементы (рис. 1$a,b$). Благодаря каналам шунтировки одного или обоих эмиттеров, или же пробою $p'n^+$ - и $p^+n'$ - переходов при обратном смещении РВД допускает пропускание обратного тока.

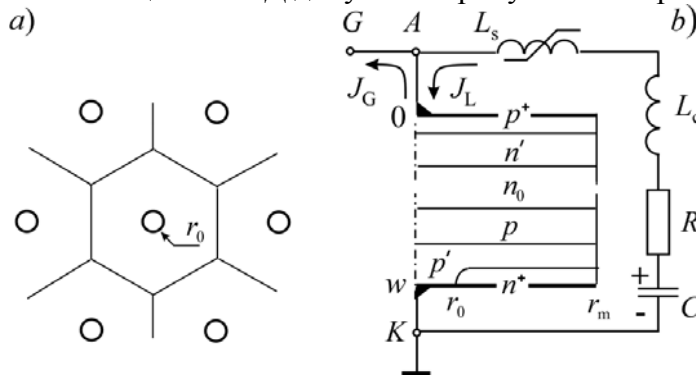

**Рис. 1.** Типовое расположение шунтированных ЭЯ по площади РВД ($a$ – вид со стороны катода); вертикальный разрез единичной ЭЯ и принципиальная рабочая схема ($b$). Здесь $A$ и $K$ – электроды анода и катода, $G$ – клемма генератора управления.

Находясь в исходном запертом состоянии в силовом контуре с заряженным до напряжения $U_{C0}$ конденсатором $C$, нагрузкой $R$, паразитной индуктивностью $L_c$ и дросселем с насыщающимся сердечником $L_s$ РВД блокирует разрядный ток. Для включения РВД к нему от генератора $G$ подается короткий управляющий импульс тока $J_G(t) < 0$. Развязка управляющего и силового контуров на этот период осуществляется за счет дросселя $L_s$, имеющего большую начальную индуктивность $L_{s0}$. На время этой развязки напряжение на дросселе резко увеличивается до значений, более высоких, чем $U_{C0}$, и он пропускает медленно растущий ток $J_L$. При достижении этим током порога, связанного с насыщением сердечника дросселя и переходом последнего в состояние с низкой индуктивностью $L_{s1} \ll L_{s0}$, направление протекания тока через прибор становится положительным и РВД осуществляет разряд конденсатора через нагрузку, в чем и состоит его функция как замыкающего ключа.

В работе [9] было обращено внимание на то, что при переходе от микро- к наносекундным временам резко снижается роль диффузии в механизме концентрационной модуляции слаболегированной толщи РВД, и восполнение этой роли может быть осуществлено только за счет сильнополевых инжекционных эффектов в условиях насыщения скоростей дрейфа свободных носителей и образования областей объемного заряда. Пространственная картина механизма включения при этом становится малодоступной для теоретического анализа, и с целью ее адекватного описания мы выбираем хорошо апробированный пакет прикладных программ фирмы Synopsys, предназначенный для имитационного компьютерного моделирования сильнонеравновесных физических процессов в различных полупроводниковых приборах [10].

## Выбор параметров моделируемой структуры и управляющих импульсов

Физические процессы во всех ЭЯ РВД протекают одинаково [1, 2], поэтому можно рассматривать весь прибор как единственную ячейку, через которую протекает $1/k$-часть тока его контактов (рис. 1). С целью единообразия интерпретации мы используем для плотностей токов в ЭЯ и токов в цепи нагрузки, а также для параметров $R$, $L$ и $C$ величины, нормированные на площадь РВД и приводимые в единицах A/cm$^2$, Ω·cm$^2$, H·cm$^2$ и F/cm$^2$, соответственно.

Предлагаемый нами компромиссный способ увеличения быстродействия РВД заключается в следующем. Вместо уменьшения полной толщины его $p$-базы до величины, меньшей 10 µm, следует оставить эту базу состоящей из двух диффузионных компонент: мелкого сильнолегированного буферного $p'$-слоя у границы с $n^+$-эмиттером толщиной $w \sim 5 - 10$ µm с типичной для тиристоров поверхностной концентрацией акцепторов $\sim 10^{18}$ cm$^{-3}$, а также сглаживающего электрические поля как в объеме, так и на защитной фаске, глубокого $p$-слоя с $w \approx 50$ µm и поверхностной концентрацией, сниженной до уровня $\sim 10^{15}$ cm$^{-3}$. При введении электронов из катодного эмиттера в $p'$-слой, обладающий сильным встроенным тянущим полем, заметного ограничения инжекции в нем не произойдет. В то же время, в глубокой слаболегированной части $p$-базы уже при умеренной мощности накачки и плотностях катодного тока порядка сотен A/cm$^2$ успеет установиться высокий уровень инжекции. Ситуация при этом станет похожей на включение РВД по квазидиодному сценарию, описанному в теоретической работе [11] и характеризуемому предельно высокой скоростью инжекционной модуляции проводимости полупроводниковой структуры.

Здесь будут рассмотрены модифицированные реверсивно-включаемые динисторы двух типов: РВД-1 на 2.5 kV с толщиной структуры 250 µm и РВД-2 на 5 kV с толщиной структуры 500 µm. Оба типа РВД имеют одинаковые ЭЯ цилиндрической формы с заданными большим радиусом $r_m = 500$ и радиусом шунта $r_0 = 15$ µm. Концентрация доноров в высокоомной $n_0$-базе равна $N_{d0} = 2 \cdot 10^{13}$ cm$^{-3}$. Профили неоднородного диффузионного легирования со стороны катодной и анодной граней пластины задаются через поверхностные концентрации $N^s_{n^+} = 10^{20}$, $N^s_{p} = 1\cdot 10^{15}$, $N^s_{p'} = 10^{19}$, $N^s_{n'} = 2\cdot 10^{17}$, $N^s_{p^+} = 10^{19}$ cm$^{-3}$ и глубины диффузии $w_{n^+} = 3$, $w_{p'} = 8$, $w_p = 50$ µm, $w_{n'} = 12$, $w_{p^+} = 4$ µm. Времена жизни электронов $\tau_{n0} = 16$ µs и дырок $\tau_{n0} = 8$ µs.

Форма импульса генератора управления задается как комбинация экспонент: $J_G(0 < t < t_s) = J_G^- - J_G^+$, $J_G^- = J_{Gm}^-[1 - \exp(-t/\tau_G^-)]$, $J_G^+ = J_{Gm}^+[1 - \exp(-t/\tau_G^+)]$, где $J_G^-$ и $J_G^-$ являются параметрами, вводимыми для учета задержки и установления амплитуды импульса $J_G(t)$. Времена задержки $\tau_G^-$ и установления $\tau_G^+$, а также коэффициенты $J_{Gm}^-$ и $J_{Gm}^+$ подбираются эмпирически. Ток через РВД вычисляется по формуле $J_D = J_L + J_G$, где $J_L$ – ток цепи нагрузки, а величина заряда $Q_R$, накапливаемого в РВД на «реверсивной» стадии ($J_D < 0$), находится как интеграл $\int J_D dt$. Температура $T = 400$ K.

## Моделирование субмикросекундных инжекционных процессов в структурах РВД и их выходные характеристики

Ориентируясь, главным образом, на частотно-импульсные приложения и полагая, что предлагаемый РВД сохранит свои ключевые показатели достаточно высокими, мы используем при задании параметров элементов силовой цепи известный принцип оптимизации быстрозатухающих колебательных режимов разряда $RLC$-контура с идеальным замыкающим ключом [12]. Его сущность состоит в достижении максимальной величины энергии, передаваемой из конденсатора в нагрузку к моменту реализации максимума на кривой $J(t)$. Для этого характеристическое сопротивление колебательного контура и сопротивление нагрузки должны соотносится в определенной пропорции $\rho = \sqrt{L/C} = 1.13R$. В этом приближении для случая РВД-1 при заданном $R = 1\,\Omega \cdot cm^2$, а также $L = 50$, $L_{s0} = 5000$ и $L_{s1} = 50\,nH \cdot cm^2$ (т.е., при суммарной индуктивности цепи нагрузки после переключения дросселя $L_\Sigma = L + L_{s1} = 100\,nH$) оптимальной для $C$ является величина $78.3\,nF$. При заданном напряжении $U_{C0} = 2.5\,kV$ это дает для начальной энергии конденсатора значение $W_{C0} = C \cdot (U_{C0})^2/2 = 0.245$ J. Величину заряда реверсивной накачки мы выбирали из соображений достаточного уменьшения переходного всплеска напряжения, и для иллюстрируемого случая она задана равной $Q_R = 4.68\,\mu C/cm^2$.

Коммутационные характеристики РВД-1 в такой цепи представлены на рис. 2. Для них введена следующая нумерация: *1* – ток через РВД, *2* – ток через цепь нагрузки, *3* – мгновенное напряжение на аноде РВД, *4* – мгновенное напряжение на конденсаторе. Кривые *5 – 7* соответствуют переменным $w_R$, $w_D$ и $w_C$, проектируемым на правую ось. Эти переменные нормированы на начальную энергию $W_{C0}$ и имеют следующий физический смысл: *5* – энергия, передаваемая в нагрузку $w_R(t) = W_R(t)/W_{C0}$, *6* – джоулевы потери в РВД $w_D(t) = W_D(t)/W_{C0}$, *7* – остаточная энергия конденсатора $w_C(t) = W_C(t)/W_{C0}$.

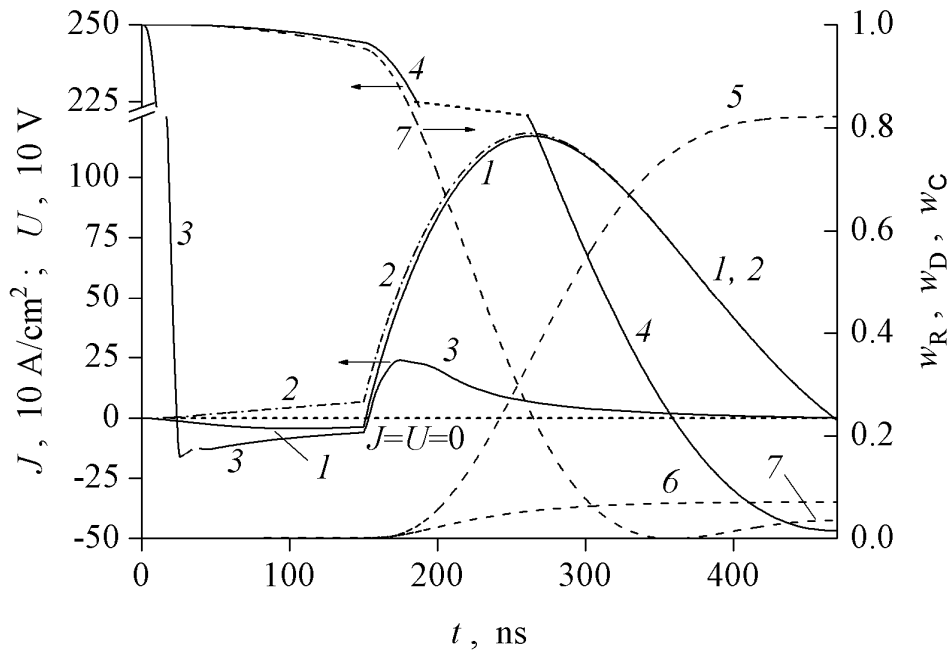

**Рис. 2.** Переходные характеристики РВД-1 (объяснения в тексте).

Обращают на себя внимание отсутствие сколь-нибудь заметной задержки нарастания тока и относительно небольшая высота всплеска напряжения $U_m = 250$ V в период формирования фронта включения. Длительность этого фронта по уровням 0.1/0.9 составляет 75 μs. Скорость нарастания тока на фронте здесь достигает 2.9 kA cm$^{-2}$s$^{-1}$. Максимальная плотность тока в импульсе составляет 1.2 kA/cm$^2$. Полная длительность коммутируемого импульса равна 320 ns.

Как видно по насыщению кривой *5*, к моменту окончания коммутируемого импульса тока при $t$ = 470 ns в нагрузку передается около 85 % начальной энергии конденсатора $W_{C0}$. Величина джоулевых потерь в РВД к этому моменту составляет 8% (кривая *6*), а величина остаточной энергии конденсатора, перезаряженного к концу импульса до отрицательного смещения $U_C \approx 0.2\,U_{C0}$, близка к 4% от величины $W_{C0}$ (кривая *7*). По абсолютным значениям это дает $W_D = 0.208$, $W_R = 0.02$ и $W_C = 0.01$ J. При частотно-импульсной работе РВД в таких режимах со средней за период мощностью охладителя 200 W/cm$^2$ ожидаемая частота повторения импульсов приближается к 1 kHz.

Для расчета случая с РВД-2 с напряжением переключения 5 kV мы сохраняем величину $R = 1\,\Omega\cdot\text{cm}^2$, и в два раза увеличиваем все индуктивные параметры, а также емкость конденсатора: $L = 100$, $L_{s0} = 10000$, $L_{s1} = 100$ nH/cm² и $C = 156.6$ nF/cm². Теперь при $U_{C0} = 5$ kV начальная энергия конденсатора увеличивается в 8 раз и составляет $W_{C0} = 1.957$ J. Заряд реверсивной накачки для этого случая выбирается из тех же соображений, что и для РВД-1, и задается равным $Q_R = 6\,\mu\text{C}/\text{cm}^2$.

Соответствующие переходные характеристики РВД-2 представлены на рис. 3. Здесь использована та же нумерация кривых, что и на рис. 2. Моменты времени, для которых далее на рис. 4 будут построены профили концентрации электронов $n(x,t)$ и поля $E(x,t)$, обозначены точками $M_1 - M_5$. Видим, что временной ход всех переменных в качественном отношении мало чем отличается от предыдущего случая. Однако, имеют место существенные количественные изменения. Амплитудная плотность коммутируемого импульса тока теперь увеличивается до 2.4 kA/cm² при длительности 650 ns, а также в два раза удлиняется фронт нарастания тока, достигая 150 ns. Высота пика напряжения при этом увеличивается до 1.75 kV, а скорость нарастания тока уменьшается до $dJ/dt = 1.75\cdot 10^{10}$ A·cm⁻²s⁻¹, что примерно в 2 раза ниже, чем в предыдущем случае, но еще остается на порядок выше типовых значений этого параметра для микросекундных режимов.

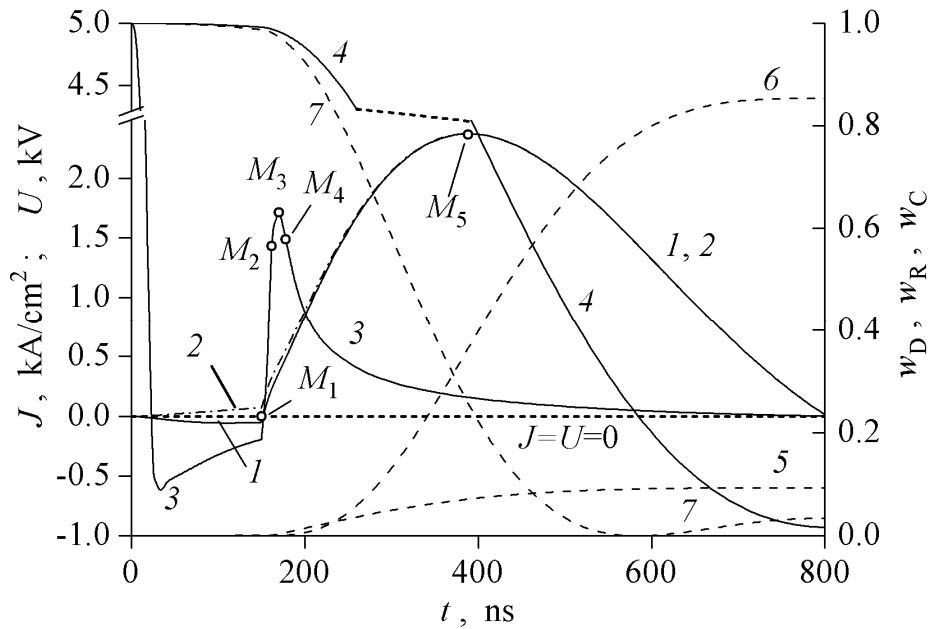

**Рис. 3**. Переходные характеристики РВД-2 (объяснения в тексте).

Что касается перераспределения начальной энергии конденсатора к завершению коммутируемого импульса, то соответствующие нормированные значения здесь равны $w_R = 0.86$, $w_D = 0.1$ и $w_C = 0.05$. При этом абсолютные значения каждой из них становятся примерно в 8 раз большими, чем для РВД-1, а именно, энергия, передаваемой в нагрузку $W_R = 1.66$, энергия потерь на нагрев РВД $W_D = 0.2$ и остаточная энергия конденсатора $W_C = 0.06$ J/cm². Предельная частота при мощности охладителя 200 W/cm² теперь должна снизиться до 100 Hz.

На рис. 4 показаны профили распределений $n(x,t)$ (сплошные кривые *1–5*) и $E(x,t)$ (штрихи *3' – 5'*) в структуре РВД-2 для следующих моментов времени *1* – 150 (точка $M_1$ на рис. 3); *2* – 162 ($M_2$); *3, 3'* – 170 ($M_3$); *4, 4'* – 178 ($M_4$); *5, 5'* – 389 ($M_5$) ns. Профили $E(x)$ для моментов $M_1$ и $M_2$ неразличимы в масштабе рисунка. Видим, что в момент $M_1$ (кривая *1*), когда начинается рабочий импульс тока, всюду в слаболегированных $n_0$- и *p*- слоях имеет место высокий уровень инжекции. На соответствующем профиле концентрации электронов выделяются 2 сильно обогащенных слоя избыточной плазмы у внешних границ слаболегированной области. Характерные концентрации $n(x,t)$ такого слоя со стороны анода составляют $\sim 5 \cdot 10^{15}$, а со стороны катода $\sim 10^{16}$ cm$^{-3}$. В период $150 < t < 170$ ns эти слои частично истощаются. Однако, примерно при $t = 178$ ns (кривая *4*) эта тенденция меняется и начинается монотонное увеличение концентраций почти всюду на слаболегированных $p$- и $n_0$-участках. После этого к моменту $t = 389$ ns, когда ток в цепи нагрузки достигает своего максимума по времени $J_m \approx 2.4$ kA/cm$^2$, концентрация плазмы в минимуме поднимается до уровня $4 \cdot 10^{15}$ cm$^2$, а поле здесь успевает снизиться до $5 \cdot 10^4$ V/cm.

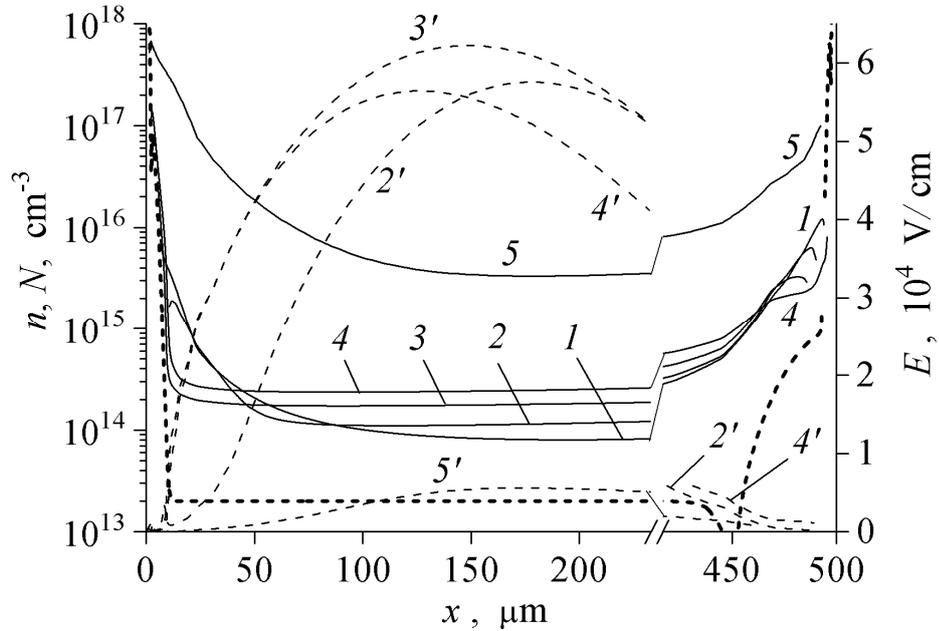

**Рис. 4**. Форма профилей $n(x,t)$ и $E(x,t)$ для РВД-2 для отдельных моментов времени. Жирной пунктирной кривой показан профиль легирования.

Важно то, что характерные величины поля в плоскостях, граничащих с плазменными слоями, на всем протяжении коммутационного процесса остаются существенно меньшими максимальной величины $E(x)$. Все эти признаки указывают на реализацию квазидиодного режима в РВД по определению в [11].

Проанализируем теперь особенности перестроек профилей $n(x,t)$ и скорости лавинной генерации $G_{aval}(x,t)$ при резком нарушении условий квазидиодного включения, чего следует ожидать при снижении включающего заряда $Q_R$ до единиц μC/cm$^2$ [9]. Для расчетов, представленных на рис. 5, значение $Q_R$ было задано равным 1.8 μC/cm$^2$. Вызывает интерес демонстрируемая последовательностью *1 – 5* волнообразная трансформация профилей $n(x,t)$ в период задержки пролета инжектируемых электронов через ту часть $p$-базы, куда не проникает поле коллектора. На вставке показано, как в этот период на кривой $U(t)$ возникают осцилляции напряжения с несколькими максимумами и минимумами. Уже через 7 ns после начала роста прямого тока к моменту времени $t$=157 ns (точка $M_1$ в первом максимуме) концентрация электронов в интервале 470 – 480 μm левее буферного слоя $p'$ (см. на рис. 4)) резко снижается. В результате между изначально накопленной плазмой и этим слоем обеднения образуется крутая ступенька высотой почти в один порядок по концентрации. Вскоре, к моменту $t$=159 ns (точка $M_2$) эта ступенька достигает геометрической плоскости коллектора (при x=450 μm , рис. 4), а затем, частично размываясь по координате, занимает позицию $\sim 350$ μm (точка $M_3$). Кривая $U(t)$ в этом интервале проходит через первый минимум, а затем за время 4 ns через промежуточное состояние $M_4$ достигается новый и самый высокий максимум (точка $M_5$) при $t$=165 ns.

На соответствующем профиле (5) видно, что в этот момент ступенька достигает окрестности буферного слоя $n'$ со стороны анода. В это же время, на участке 250 – 450 μm становится заметным рост скорости ударной генерации $G_{aval}(x,t)$ (кривая 6). Как видно по профилю $n(x,t)$ на этом участке в момент $M_5$, тенденция на истощение концентрации электронов меняется в сторону их накопления, несмотря на то, что поступление новых электронов их катода еще остается ограниченным.

В последующий период (правее точки $M_5$) напряжение на РВД быстро снижается и лавинный пробой должен прекратиться. Тем не менее, на кривой $U(t)$ реализуется еще один максимум. Это можно объяснить, как результат действия эффектов, аналогичных тем, что наблюдаются в инжекционно-пролетных диодах [13], но теперь протекающих в условиях со значительно большей абсолютной коммутируемой мощностью и при более сложном взаимодействии с цепью нагрузки.

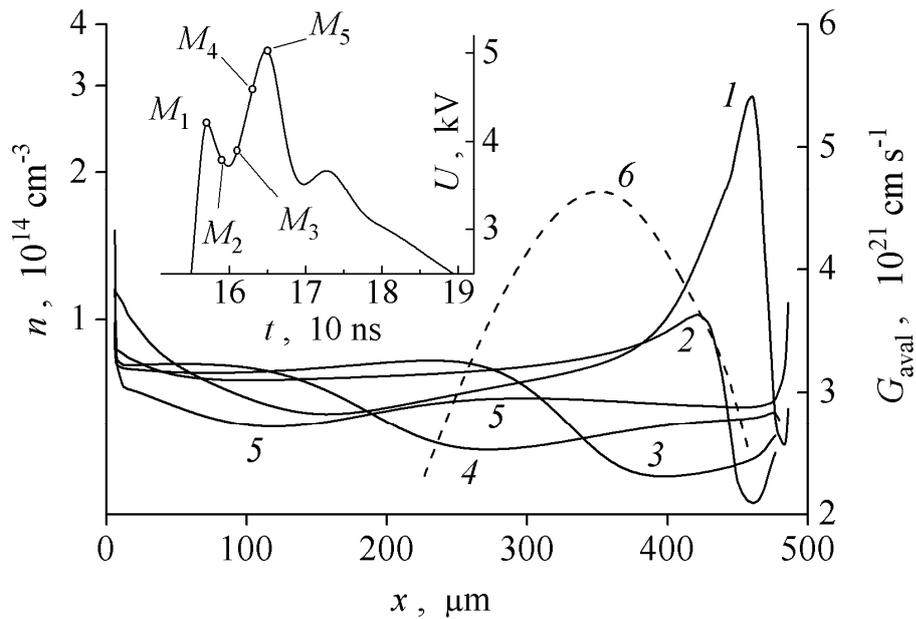

**Рис. 5.** Профили концентрации электронов $n(x,t)$ (сплошные кривые 1–5) и скорости лавинной генерации $G_{aval}(x,t)$ (кривая 6) в моменты времени 1 – 157, 2 – 159, 3 – 161, 4 – 163, 5 – 165 и 6 – 165 μs. На вставке – форма всплеска напряжения в окрестности его вершины.

Последующий переход к стадии монотонного уменьшения напряжения говорит об установлении регенеративной обратной связи по току, однако нельзя не заметить, что это происходит после крайне нежелательной задержки около 30 ns. Как известно из физических основ надежности мощных полупроводниковых приборов [14], подобные волнообразные процессы инжекции и осцилляции напряжения и тока весьма характерны при их электрической перегрузке. Более того, в переключателях с распределенным импульсным управлением и с большой рабочей площадью таких как РВД, а также в силовых интегральных микросхемах [15], они могут сильно осложняться из-за пространственной неустойчивости тока, инициируемой небольшими технологическими флуктуациями или конструкционной неэквивалентностью отдельных групп управляемых ячеек, и способной развиваться в течение десятков ns в аварийные сценарии локализации тока и тепла.

Далее, на рис. 6 приведены зависимости некоторых важных выходных параметров РВД-1 (1 – 4) и РВД-2 (1' – 4') от величины включающего заряда $Q_R$. На левую ось спроектированы кривые, отражающие зависимости от $Q_R$ величин $w_R$ (1, 1'), $w_R$ (2, 2') и $w_R$ (3, 3') в момент окончания коммутируемого импульса. Кривыми 4, 4' представлены зна-чения поля $E_m = E(x_{Em}, r_m, t_{Em})$ в точке его максимума по координате $x_{Em}$, и времени $t_{Em}$ (правая ось).

Видим, что при достаточно большом увеличении $Q_R$, когда в РВД реализуются условия квазидиодного режима и они по своим характеристикам приближаются к идеальным ключам, происходит постепенное насыщение всех передаточных энергетических переменных, причем величины $w_D$ и $w_R$ становятся при $Q_R > 4.5$ μC/cm² примерно одинаковыми. Но при зарядах накачки, меньших 2.5 – 4.0 μC/cm², различия усиливаются, что особенно заметно для $w_D$ и $w_R$. Действительно, при слабых управляющих импульсах из-за увеличения выбросов напряжений и задержек тока растут потери на нагрев структур РВД и тем самым отнимается доля полезной энергии, передаваемой в нагрузку. При этом по своим коммутационным характеристикам РВД начинают удаляться от идеального ключа.

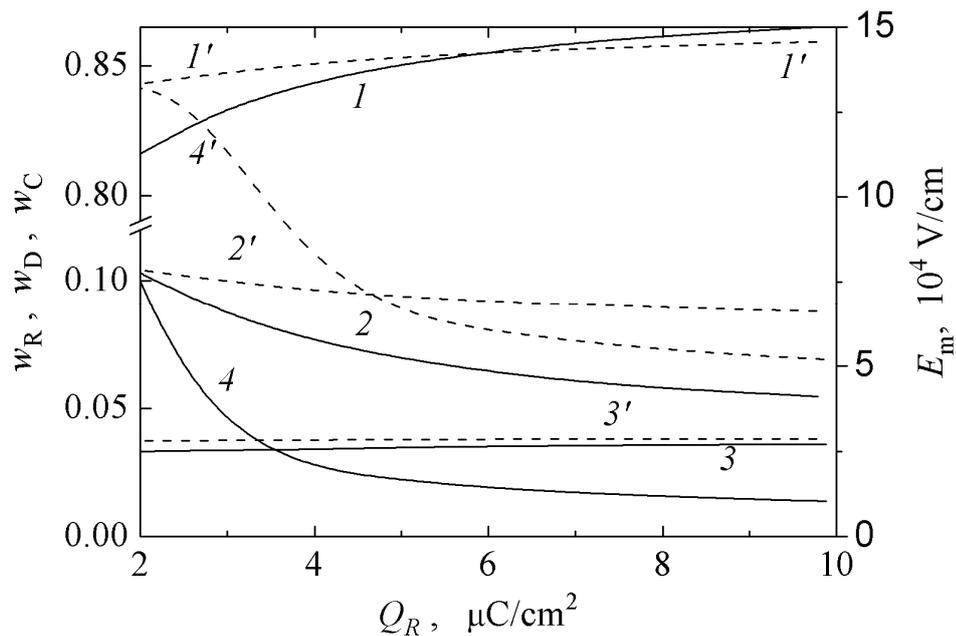

**Рис. 6.** Зависимость некоторых параметров предлагаемых модификаций РВД от $Q_R$. Семейства кривых *1 – 4* относятся к РВД-1, а кривых *1' – 4'* к РВД-2 (объяснения в тексте).

Заметим, что соответствующие режимы существенно различаются по максимальной величине поля $E_m$ в структурах РВД-1 и РВД-2. Так, в РВД-1 во всем интервале изменения $Q_R$ эта величина с большим запасом не превышает порога электрического пробоя. Однако в РВД-2 при снижении $Q_R$ до 2 μC/cm² и ниже величина поля проходит через некоторый условный рубеж $1.25 \cdot 10^5$ V/cm, после чего начинает насыщаться. Это свидетельствует о начале лавинного пробоя в окрестностях плоскости $x = x_{Em}$.

**Заключение**

Имитационное моделирование реверсивно-включаемых динисторов подтверждает возможность их использования для коммутации мощных импульсов субмикросекундного диапазона при условии, что будут удовлетворены некоторые обязательные требования к их конструкции. В частности, для устранения опасной задержки и чрезмерных джоулевых потерь на этапе нарастания тока при коммутации с помощью РВД необходимо, чтобы была снижена поверхностная концентрация глубокого акцепторного слоя в *p*-базе до величины ~ $10^{15}$ cm⁻³. При выполнении этого требования, как показывают результаты имитационного эксперимента, можно рассчитывать на существенное увеличение скорости на-растания тока на фронте включения до $(2 – 3)10^{10}$ A·cm$^{-2}$s$^{-1}$ и сокращение длительности этого фронта до 75 – 150 ns при максимальной амплитудной плотности тока в единицы kA/cm$^2$ и при вполне достижимых зарядах реверсивной накачки ~ 4 – 6 μQ/cm$^2$. Конечно, параметры дросселя должны быть такими, чтобы его переход в состояние с малой индуктивностью в каждом конкретном случае происходил не раньше, чем будет накоплен достаточный для эффективного запуска РВД управляющий заряд.

В интервале начальных напряжений на силовом емкостном накопителе от 2.5 до 5 киловольт и при оптимальном сочетании параметров элементов силовой цепи в нагрузку может быть передано от 0.2 до 1.7 J начальной энергии накопителя. При этом в условиях стандартного охлаждения ~ 200 W/cm$^2$ ожидаемая частота повторения импульсов будет лежать в интервале 100 – 1000 Hz. Реверсивно-включаемые динисторы с такими характеристиками смогут найти широкое применение в современной импульсной энергетике.